\documentstyle[prb,aps,multicol]{revtex}  
\begin{document}                
\draft
\title{Electronic structure of In$_{1-x}$Mn$_x$As
studied by photoemission spectroscopy: Comparison with Ga$_{1-x}$Mn$_x$As}

\author{J. Okabayashi, T. Mizokawa, D. D. Sarma$^{\ast}$, and A. Fujimori}
\address{Department of Complexity Science and Engineering and
  Department of Physics, University of Tokyo Bunkyou-ku, Tokyo
  113-0033, Japan}

\author{T. Slupinski$^{**}$, A. Oiwa, and H. Munekata$^{***}$}
\address{Kanagawa Academy of Science and Technology (KAST), 
3-2-1 Sakado, Takatsu-ku, Kawasaki, 213-0012, Japan} 

\date{Received {\today}}
\maketitle
\begin{abstract}
 We have investigated the electronic structure of the $p$-type diluted
 magnetic semiconductor In$_{1-x}$Mn$_x$As by photoemission
 spectroscopy. The Mn 3$d$ partial density of states is found to be basically 
 similar to that of 
 Ga$_{1-x}$Mn$_x$As. However, the impurity-band like states near the
 top of the valence band have not been observed by angle-resolved
 photoemission spectroscopy unlike Ga$_{1-x}$Mn$_x$As. This
 difference would explain the
 difference in transport, magnetic and optical properties of 
 In$_{1-x}$Mn$_x$As and Ga$_{1-x}$Mn$_x$As. The different electronic
 structures are attributed to the weaker Mn 3$d$ - As 4$p$
 hybridization in In$_{1-x}$Mn$_x$As than in Ga$_{1-x}$Mn$_x$As.  
\end{abstract}
\begin{multicols}{2}

\narrowtext
Diluted magnetic semiconductors (DMS) have attracted much attention
because of the combination of magnetic and semiconducting properties
and hence high potential for new device applications. Recently DMS
based on  III-V compounds have been extensively studied because of
the success in doping high concentrations of transition-metal ions by
molecular beam epitaxy (MBE).\cite{Munekata89} Most remarkably, Mn 
doping in InAs and GaAs leads to 
ferromagnetism and interesting magneto-transport properties.
\cite{Ohno_Science,Iye99} This behavior is
generally called ``carrier-induced ferromagnetism'' because hole
carriers introduced into the system mediate the ferromagnetic coupling 
between the Mn ions \cite{OhnoMMM} although its
microscopic mechanism has been controversial until
now. The key to clarify the mechanism of the carrier-induced 
ferromagnetism is to
understand the nature of the doped hole carriers as well as the
exchange interaction between the holes in host valence band and the
localized $d$ orbitals of the magnetic ions, so-called $p$-$d$ 
exchange interaction.
For Ga$_{1-x}$Mn$_x$As, previous investigations including
photoemission studies \cite{Jun99,Jun01} have revealed that the
basically localized Mn 3$d$ electrons interact with the
doped holes through the $p$-$d$
hybridization which causes the $p$-$d$ exchange
interaction.\cite{OhnoMMM} 

As for the closely related system In$_{1-x}$Mn$_x$As\cite{Slupinski}, 
there are several differences from 
Ga$_{1-x}$Mn$_x$As.\cite{OhnoMMM} (1) The Curie temperature is
relatively low:
$T_c{\leq}45$ K. (2) Optical absorption measurements have
indicated that
In$_{1-x}$Mn$_x$As shows a Drude-like behavior due to free
carriers.\cite{Hirakawa}  For Ga$_{1-x}$Mn$_x$As, on the other hand,
a broad peak was observed around 200 meV and there was no clear Drude
component.\cite{HirakawaGMA} 
So far, no photoemission study 
has been reported for In$_{1-x}$Mn$_x$As. 
Comparative studies of 
Ga$_{1-x}$Mn$_x$As and In$_{1-x}$Mn$_x$As would provide us with useful 
information to elucidate the origin of the ferromagnetism in these systems. 
The electronic structure of In$_{1-x}$Mn$_x$As itself is also
interesting because it has been reported that In$_{1-x}$Mn$_x$As grown 
on GaSb substrate exhibits 
photo-carrier induced ferromagnetism,\cite{Koshihara,Oiwa} and
field-induced ferromagnetism.\cite{Ohno_nature}  
Investigation of the electronic structure, especially of the
hybridization and the exchange 
interaction between the host valence band and the 
localized magnetic ions, as well as the magnetic coupling between the 
magnetic ions, would give us a useful guideline for further 
development in functional materials design. Although several
theoretical models have been
discussed\cite{Akai,Inoue,Dietil,Kanamori}, there are few
experimental investigations to clarify the electronic structure and the 
mechanism of carrier-induced ferromagnetism. 
The purpose of this paper is to clarify the 
electronic structure of In$_{1-x}$Mn$_x$As by resonant photoemission 
spectroscopy (RPES), which yields the Mn 3$d$ partial density of 
states (DOS), and by angle-resolved
photoemission spectroscopy (ARPES), which measures the 
energy-band dispersions. We compare the present results with
those of Ga$_{1-x}$Mn$_x$As.\cite{Jun99,Jun01} 

A $p$-type In$_{1-x}$Mn$_x$As/GaSb sample of 30 nm thickness was grown by
 MBE. \cite{Munekata93,Slupinski} 
The sample had a Curie temperature of $35$ K and the Mn content
was estimated to be $x$=0.09 based on the calibration of beam fluxes during 
MBE growth. Experiments were 
performed at beamline BL 18-A of Photon Factory, High Energy Accelerator
Research Organization, using an ADES-500 analyzer for ARPES and a
CLAM-II analyzer for angle-integrated RPES. The 
total energy resolution was set to 100 meV, comparable to the thermal
broadening at the room temperature where all experiments were carried
out. The angular resolution of the ADES-500 and CLAM-II analyzers were
${\pm}$1$^{\circ}$, ${\pm}$4$^{\circ}$, respectively. To remove oxidized surface layers and other 
contamination, we made repeated Ar-ion sputtering (1 kV) and
annealing. The annealing temperature was limited to $200^{\circ}$C to 
avoid the segregation of MnAs clusters.\cite{Katsumoto} The 
cleaned surface showed a 1${\times}$1 LEED pattern. We checked the
chemical composition of the sample by measuring x-ray photoemission 
spectra of In, As, and Mn core levels. 
For angle-integrated RPES in the Mn 3$p$ to 3$d$ absorption region, 
photons of $h{\nu}=46-55$ eV were used. 
In the case of ARPES,
electrons emitted in the direction normal to the surface, which come from
the ${\Gamma}-{\Delta}-X$ line in the Brillouin zone, were collected. 
We compared the In$_{1-x}$Mn$_x$As spectra with those of the reference 
$p$-type InAs. The identical procedure of surface cleaning 
was performed also for the InAs sample. Binding energies are referenced
to the Fermi edge of Ta spectra in electrical contact with the sample.

Figure 1 shows RPES spectra recorded using photon energies of 
$h{\nu}$=$46-55$ eV. The spectra have been normalized to the photon flux.
Resonant enhancement occurred at $h{\nu}$=50 eV and
off-resonance spectra were taken at $h{\nu}$=48 eV. 
The difference between the two curves yields the Mn 
3$d$-derived spectra as in the case of Ga$_{1-x}$Mn$_x$As.\cite{Jun99}  
The difference spectrum shows a sharp peak at ${\sim}$4 eV 
binding energy as well as a broad peak at ${\sim}$7 eV binding
energy. There is little intensity at the Fermi level ($E_F$). Such a spectral 
line shape is almost identical to that of Ga$_{1-x}$Mn$_x$As, which has been interpreted in terms of
configuration-interaction cluster-model calculations. From that analysis,
the ground state 
has been found to be dominated by the 3$d^5$ configuration.\cite{Jun99} 
These spectra also indicate strong hybridization 
between the Mn 3$d$ electrons and the As 4$p$-derived valence band.

As shown at the bottom of Fig. 1, we compare the experimental Mn 
3$d$ partial DOS of In$_{1-x}$Mn$_x$As with 
configuration-interaction cluster-model calculations. For comparison,
we also show the result for Ga$_{1-x}$Mn$_x$As. 
The parameters for the cluster-model calculations are the 
charge-transfer energy ${\Delta}$ from the ligand $p$ level to the 
transition-metal $d$ level, the on-site Coulomb 
energy $U$ between two Mn 3$d$
electrons, and the hybridization strength $(pd{\sigma})$ between the 
ligand $p$ orbital and the transition-metal $d$ orbitals defined by 
Slater-Koster parameters. The details of the calculations are given in 
Ref. \onlinecite{Mizokawa}. The Mn 3$d$-derived spectrum has been 
reproduced using parameters with a smaller ($pd{\sigma}$) value compared 
to that for Ga$_{1-x}$Mn$_x$As as
summarized in Table I.
According to the cluster-model
calculations, the ${\sim}$2 eV feature predominantly consists of 
$d^5\underline{L}$ final states, while the satellite 
structure comes from the $d^4$ final-state configuration because 
$E(d^4)-E(d^5\underline{L}){\sim}U-{\Delta}>0$. 
Using the above parameters, we have estimated the 
$p$-$d$ exchange 
interaction in In$_{1-x}$Mn$_x$As to be $N{\beta}=-0.7$ eV. This 
value is smaller than that of 
Ga$_{1-x}$Mn$_x$As ($N{\beta}=-1.0$ eV).\cite{Jun99}
The deppressed intensity 
at $E_F$ in the difference spectrum compared to the valence-band
intensity at $E_F$ for various photon energies both for
In$_{1-x}$Mn$_x$As and Ga$_{1-x}$Mn$_x$As indicates
that the Mn 3$d$ DOS are not dominant at $E_F$ and
suggests that 
hole carriers of As 4$p$ character contribute to the transport.

The difference spectrum between In$_{1-x}$Mn$_x$As and pure InAs taken 
at $h{\nu}=70$ eV has also been used to obtain the Mn 3$d$ partial 
DOS as shown in Fig. 2. The spectra have been normalized to 
the In 4$d$ core-level peaks taking into account the composition
difference between In and Mn. Because of the Cooper minimum of the As
4$p$ states at $h{\nu}{\sim}70$ eV, the ionization cross section of As 4$p$
reaches a minimum, and the Mn 3$d$ component is relatively 
enhanced.\cite{Yah} The
lineshape of the difference spectrum is almost the same as that in Fig. 1, 
which guarantees that the Mn 3$d$ density of states deduced from RPES 
is valid to discuss the electronic structure.   

To obtain more information about the electronic structure around $E_F$, we
have measured ARPES spectra along the
${\Gamma}-{\Delta}-X$ line for both 
In$_{1-x}$Mn$_x$As and InAs. As shown in Fig. 3, by changing the
photon energy in the normal 
emission set up from the (001) surface \cite{Hufner}, the energy band 
dispersions along the
${\Gamma}-{\Delta}-X$ line were observed. According to the
direct-transition model for ARPES with an appropriate inner 
potential (10 eV)\cite{Aono}, the ${\Gamma}$ point is measured at $h{\nu}{\sim}10$ eV
and the $X$ point is measured at $h{\nu}{\sim}32$ eV. The peaks in Fig. 3
(a) correspond to the ${\Delta}_1$ band (split-off band) and
${\Delta}_3+{\Delta}_4$ band (heavy- and light-hole bands,
respectively) along the ${\Gamma}-{\Delta}-X$ line. The clear dispersion 
curves are almost the same as those for GaAs 
and InAs. Due to the Mn 3$p$-3$d$ resonant 
effect, the line shape changes drastically around $h{\nu}=50$ eV, 
almost in the same way as in the RPES result.  
Comparison of the spectra near $E_F$ between
In$_{1-x}$Mn$_x$As and InAs shown in Fig. 3(b) reveals no clear
differences in spite of Mn doping. This is quite different from the case of 
Ga$_{1-x}$Mn$_x$As and GaAs, where impurity-band-like new states were 
found to form near the 
valence-band maximum (VBM) by Mn doping as shown in 
Fig. 3(c). Although Mn doping in GaAs
induces split-off states above the VBM through hybridization with As
4$p$ and forms the impurity band-like states, Mn in InAs does not induce
such states, probably because of the weaker hybridization strength
($pd{\sigma}$) than in GaAs. 
In fact, in the dilute limit of Mn in InAs (Mn: 8${\times}10^{16}$ cm$^{-3}$), 
Mn doping leads to the formation of an
acceptor level of primarily As 4$p$ character at 30 meV above the 
VBM,\cite{acceptor} which is much smaller than that
for Mn in GaAs (100 meV).\cite{GMA_acceptor}

Now we discuss the origin of the differences between the electronic
structure of In$_{1-x}$Mn$_x$As and Ga$_{1-x}$Mn$_x$As. 
Comparing Mn 3$d$ spectral features in In$_{1-x}$Mn$_x$As and
Ga$_{1-x}$Mn$_x$As, we find that the main peak in In$_{1-x}$Mn$_x$As
has a lower binding energy compared to that in Ga$_{1-x}$Mn$_x$As,
suggesting a lower ${\Delta}$ in In$_{1-x}$Mn$_x$As. 
There are two possible reasons for the decrease of the hybridization strength 
($pd{\sigma}$) in In$_{1-x}$Mn$_x$As compared to that in 
Ga$_{1-x}$Mn$_x$As. 
One is the differences in the Mn-As distance in these two systems. 
Extended x-ray-absorption fine-structure (EXAFS)
measurements were performed and the Mn-As distance was reported to 
be 2.44 \AA~ for Ga$_{1-x}$Mn$_x$As\cite{Shioda}
and 2.54-2.58 \AA~ for In$_{1-x}$Mn$_x$As.\cite{Soo} The longer 
Mn-As distance in In$_{1-x}$Mn$_x$As would lead to a decrease in the $p$-$d$
hybridization strength. 
Second, because of the smaller band gap in InAs (0.4 eV) than that in GaAs
(1.5 eV), the In 5$s$
states in the unoccupied part is more strongly mixed with the As 4$p$ 
states in the valence band than in the case of Ga 4$s$ states in
GaAs. Through the 
mixing with the In 5$s$ states, the As 4$p$ weight in the valence
band is reduced and hence the hybridization 
between the valence band and the Mn 3$d$ state would decrease. 
Due to the weaker hybridization between the host valence band and Mn
3$d$, the acceptor levels in In$_{1-x}$Mn$_x$As are not so strongly
split off from the valence band maximum (VBM) and the Mn ions become
difficult to bind holes. The bound hole picture 
(Mn$^{2+}$ + bound hole) is therefore less
appropriate for In$_{1-x}$Mn$_x$As than for Ga$_{1-x}$Mn$_x$As, and 
holes in In$_{1-x}$Mn$_x$As behave as free carriers. 
This would naturally explain the difference in the optical properties 
of In$_{1-x}$Mn$_x$As and
Ga$_{1-x}$Mn$_x$As.\cite{Hirakawa,HirakawaGMA}

The formation of the impurity-band-like states around $E_F$ would be 
important for the magneto-transport properties of the Mn doped DMS. 
According to the 
electron paramagnetic resonance (EPR) measurements,  
the Mn 3$d$ signals of Ga$_{1-x}$Mn$_x$As\cite{EPRGMA} showed 
that the Mn impurities in this system were predominantly 
in the ionized
state (Mn$^{2+}$, $A^-$). Similar picture was also obtained for
$n$-type In$_{1-x}$Mn$_x$As\cite{EPRIMA} in which incorporated Mn in
ionized by the excess donors. From
the view point of photoemission spectroscopy,
the Mn 3$d$ spectrum in Fig. 1 is analyzed using of the cluster model
with the Mn$^{2+}$ ground state as in the case of 
Ga$_{1-x}$Mn$_x$As. These results support that the Mn 3$d$
electronic configuration is similar between Ga$_{1-x}$Mn$_x$As and 
In$_{1-x}$Mn$_x$As. However, it is difficult to distinguish between
the Mn$^{2+}$
states with a weakly bound hole and with a free hole. In
both In$_{1-x}$Mn$_x$As and Ga$_{1-x}$Mn$_x$As, the itinerant holes 
obviously mediate the ferromagnetic order.\cite{OhnoMMM}  The ferromagnetism in the 
double-perovskite 
compound Sr$_2$FeMoO$_6$ is also considered to be caused by the same 
mechanism as the Mn-doped DMS in the sense that the O 2$p$ doped hole 
mediates the ferromagnetism through the gain in kinetic energy in the 
ferromagnetic state.\cite{Sarma,Kanamori}
The peculiar feature in Ga$_{1-x}$Mn$_x$As is that impurity-band-like
states are split off from the VBM, suggesting virtually bound holes rather
than simple free carriers of In$_{1-x}$Mn$_x$As. The formation of 
these states may help to increase the $T_c$ of Ga$_{1-x}$Mn$_x$As 
compared to In$_{1-x}$Mn$_x$As but more studies are necessary to
clarify the mechanism for this.

In conclusion, we have investigated the electronic structure of 
In$_{1-x}$Mn$_x$As using ARPES and RPES and 
compared it with Ga$_{1-x}$Mn$_x$As. 
Although the Mn 3$d$ states are in the Mn$^{2+}$
configuration in both systems, impurity-band-like 
states are not observed in In$_{1-x}$Mn$_x$As unlike
Ga$_{1-x}$Mn$_x$As. We attribute this to
the weaker hybridization in In$_{1-x}$Mn$_x$As than in
Ga$_{1-x}$Mn$_x$As, that is not sufficient
to split off states from the VBM to form the impurity-band-like states. 

This
work was performed under the approval of the Photon Factory Program 
Advisory Committee (Proposal No. 99G140). One of the authors (JO)
acknowledges support from the Japan Society for the Promotion of
Science for Young Scientists. 

\begin{table}
\caption{Electronic structure parameters $\Delta$, $U$ and
  $(pd\sigma)$ and the exchange coupling
  constant $N{\beta}$ for substitutional 
Mn impurities in InAs and GaAs in units of eV. Error bars from 
the lineshape analyses are $\pm0.5$ eV for $\Delta$, $U$
and $N{\beta}$, and $\pm0.1$ eV for $(pd\sigma)$.}
\begin{tabular}{lccccr}
 Material&${\Delta}$&$U$&($pd{\sigma}$)&$N{\beta}$& \\
\tableline
Ga$_{1-x}$Mn$_{x}$As&1.5&3.5&1.0&1.0 &Ref.\onlinecite{Jun99} \\
In$_{1-x}$Mn$_{x}$As&1.0&3.5&0.8&0.7 &this work \\
\end{tabular}
\end{table}


\begin{figure}
 \caption{A series of photoemission spectra of In$_{0.9}$Mn$_{0.1}$As
   at various photon energies in the Mn 3$p$ - 3$d$ core excitation
  threshold. The difference spectra between the on-resonant
  ($h{\nu}$=50 eV) and off-resonant (48 eV) spectra, which is a
  measure of the Mn 3$d$ partial density of states, is shown at the
  bottom and is compared with configuration interaction cluster 
  calculation results
  assuming the Mn$^{2+}$ ground state configuration. Similar
  comparison is shown for Ga$_{1-x}$Mn$_x$As (Ref. 6).}
\end{figure}

\begin{figure}
 \caption{Angle-integrated photoemission spectra of
   In$_{0.9}$Mn$_{0.1}$As and InAs at
   $h{\nu}=70$ eV, where the Mn 3$d$ cross section is large compared
   with As 4$p$, and their difference spectrum representing the Mn
   3$d$ partial density of states. }
\end{figure}

\begin{figure} 
 \caption{Angle-resolved photoemission spectra of\\ 
   In$_{1-x}$Mn$_x$As along the ${\Gamma}-{\Delta}-X$
   direction. (a) Wide-range spectra. Vertical bars show peak or 
   shoulder positions. (b) Narrow-range spectra near the Fermi level 
   for In$_{0.9}$Mn$_{0.1}$As (solid curves) and InAs (dashed
   curves). (c) Corresponding spectra for Ga$_{1-x}$Mn$_x$As and GaAs
   taken from Ref. 7.}
\end{figure}

\end{multicols}

\begin{references}

\bibitem[*]{adr} Permanent address: Solid State and Structural Chemistry Unit,
  Indian Institute of Science, Bangalore 560 012, India. 
\bibitem[**]{adr} Present address: Institute of Experimental Physics,
  Warsaw University, Hoza 69, 00-681 Warsaw, Poland.
\bibitem[***]{adr} Original post at: Imaging Science and Engineering 
  Laboratory, Tokyo Institute of Technology, Nagatsuda, Midori-ku, Yokohama
  226-8503, Japan.

\bibitem{Munekata89} H. Munekata, H. Ohno, S. von Molnar,
  A. Segmuller, and L. L. Chang, Phys. Rev. Lett., {\bf63}, 1849
  (1989).

\bibitem{Ohno_Science} H. Ohno, Science, {\bf281}, 951 (1998).

\bibitem{Iye99} A. Oiwa, A. Endo, S. Katsumoto, Y. Iye,
  H. Ohno, and H. Munekata, Phys. Rev. B, {\bf59}, 5826 (1999);
  Y. Iye, A. Oiwa, A. Endo, S. Katsumoto, F. Matsukura, A. Shen, 
  H. Ohno, and H. Munekata, Mat. Sci. Eng. B, {\bf63}, 88 (1999).

\bibitem{Slupinski} T. Slupinski, A. Oiwa, S. Yanagi, and H. Munekata, 
  J. Cryst. Growth, in press. 

\bibitem{OhnoMMM} H. Ohno, J. Mag. Mag. Matel., {\bf200}, 110 (2000).

\bibitem{Jun99} J. Okabayashi, A. Kimura, T. Mizokawa, A. Fujimori,
  T. Hayashi, and M. Tanaka, Phys. Rev. B, {\bf59}, R2846 (1999).

\bibitem{Jun01} J. Okabayashi, A. Kimura, O. Rader, T. Mizokawa,
  A. Fujimori, T. Hayashi, and M. Tanaka, Phys. Rev. B, {\bf64},
  125304 (2001).

\bibitem{Hirakawa}  K. Hirakawa, A. Oiwa, and H. Munekata, Physica E,
  {\bf10}, 215 (2001).

\bibitem{HirakawaGMA} K. Hirakawa, and S. Katsumoto, Phys. Rev. B, in
  press.

\bibitem{Koshihara} S. Koshikara, A. Oiwa, M. Hirasawa, S. Katsumoto,
  Y. Iye, C. Urano, H. Takagi, and H. Munekata, Phys. Rev. Lett.,
  {\bf78}, 4617 (1997).

\bibitem{Oiwa} A. Oiwa, T. Slupinski, and H. Munekata,
  Appl. Phys. Lett. {\bf78} 518 (2001).

\bibitem{Ohno_nature} H. Ohno, D. Chiba, F. Matsukura, T. Omiya,
  E. Abe, T. Dietl, Y. Ohno, and K. Ohtani, Nature, {\bf408},  944
  (2000). 

\bibitem{Akai} H. Akai, Phys. Rev. Lett., {\bf81}, 3002 (1998).

\bibitem{Inoue} J. Inoue, S. Tonoyama, and H. Itoh, Phys. Rev. Lett,
  {\bf85}, 4610 (2000).

\bibitem{Dietil} T. Dietl, H. Ohno, F. Matsukura, J. Cibert, and
  D. Ferrand, Science, {\bf287}, 1019 (2000). 

\bibitem{Kanamori} J. Kanamori, and K. Terakura, J. Phys. Soc. Jpn.,
  {\bf70}, 1433 (2001).

\bibitem{Munekata93} H. Munekata, A. Zaslavsky, P. Fumagalli, and
  R. J. Gambino, Appl. Phys. Lett., {\bf63}, 2929 (1993).

\bibitem{Katsumoto} S. Katsumoto, unpublished.

\bibitem{acceptor} E. I. Georgitse, I. T. Postolaki, V. A. Smirnov,
  and P. G. Untila, Sov. Phys. Semicond., {\bf23 (4)}, 469 (1989).

\bibitem{GMA_acceptor} M. Linnarsson, E. Janzen, B. Monemar, M,
  Kleverman and Thilderkvist, Phys. Rev. B {\bf55}, 6938 (1997).


\bibitem{Mizokawa} T. Mizokawa and A. Fujimori, Phys. Rev. B, 
{\bf48}, 14150 (1993).

\bibitem{Yah} J. -J. Yeh and I. Lindau, At. Data Nucl. Data Tables
  {\bf32}, 1 (1985).

\bibitem{Hufner} S. H\"{u}fner, {\it Photoelectron Spectroscopy}
  (Springer-Verlag,\\
1995). 

\bibitem{Aono} T. C. Chiang, J. A. Knapp, M. Aono, and
  D. E. Eastman,
  Phys. Rev. B {\bf21}, 3513 (1980).  

\bibitem{Shioda} R. Shioda, K. Ando, T. Hayashi, and M. Tanaka,
  Phys. Rev. B, {\bf58}, 1100 (1998).

\bibitem{Soo} Y. L. Soo, S. W. Huang, Z. H. Ming, T. H. Kao,
  H. Munekata, and L. L. Chang, Phys. Rev. B, {\bf53}, 4905 (1996). 

\bibitem{EPRGMA} J. Szczytko, A. Twadowski, K. Swiatek, M. Palczewska, 
  T. Hayashi, M. Tanaka, K. Ando, Phys. Rev. B, {\bf60}, 8304 (1999).

\bibitem{EPRIMA} J. Szczytko, A. Twadowski, M. Palczewska, R. Jalonski,
  J. Furdyna, and H. Munekata, Phys. Rev. B, {\bf63}, 085315 (2001).

\bibitem{Sarma} D. D. Sarma, P. Mahadevan, T. S. Dasgupta, S. Ray, and 
  A. Kumar, Phys. Rev. Lett., {\bf85}, 2549 (2000).: D. D. Sarma,
  Curr. Opin. in Solid State Matel. Sci., {\bf5}, 261 (2001).

\end{references}
\end{document}